\documentclass{article}
\usepackage{spconf,amsmath,graphicx,afterpage,float,booktabs,multirow}

\usepackage{enumitem}
\setlist{nosep, leftmargin=14pt}

\usepackage{mwe} 

\title{TexLiverNet: Leveraging Medical Knowledge and Spatial-Frequency Perception for Enhanced Liver Tumor Segmentation}
\name{Xiaoyan Jiang$^{1 \star \dagger}$, Zhi Zhou$^{1 \star}$, Hailing Wang$^{1}$, Guozhong Wang$^{1}$, Zhijun Fang$^{2}$}
\address{$^{1}$ School of Electronic and Electrical Engineering, \\Shanghai University of Engineering Science, Shanghai, China\\
    $^{2}$ School of Computer Science and Technology, Donghua University, Shanghai, China
}
\begin{document}
%
\maketitle
\begin{abstract}
Integrating textual data with imaging in liver tumor segmentation is essential for enhancing diagnostic accuracy.
However, current multi-modal medical datasets offer only general text annotations, lacking lesion-specific details critical for extracting nuanced features, especially for fine-grained segmentation of tumor boundaries and small lesions. 
To address these limitations, we developed datasets with lesion-specific text annotations for liver tumors and introduced the TexLiverNet model. 
TexLiverNet employs an agent-based cross-attention module that integrates text features efficiently with visual features, significantly reducing computational costs. 
Additionally, enhanced spatial and adaptive frequency domain perception is proposed to precisely delineate lesion boundaries, reduce background interference, and recover fine details in small lesions. 
Comprehensive evaluations on public and private datasets demonstrate that TexLiverNet achieves superior performance compared to current state-of-the-art methods.
Code is available at https://github.com/sky-visionX/TexLiverNet.
\end{abstract}
\section{Introduction}
\label{sec:intro}

Deep learning is widely applied to accurately map liver anatomy, supporting treatment planning for procedures like tumor resection and minimally invasive surgery. U-Net \cite{unet} and its variants \cite{unet++,attention} are especially popular for their efficient encoder-decoder structure, optimized for lesion feature extraction from medical images. However, most liver tumor segmentation methods \cite{lits,sbcnet} rely on single-image modality potentially limiting the depth of tumor characterization.

Recently, the availability of annotated text data has advanced text-guided medical image segmentation, which integrates complex medical text with image data. 
Attention-based methods have shown particular promise. For instance, TGANet \cite{tganet} translates textual descriptions into attention-weighted text features, enhancing polyp segmentation when combined with image data. LanGuideSeg \cite{ariadne}  extracts text features using CXR-BERT, and realizes deep fusion of text and image information at the decoder stage through the cross-attention mechanism, effectively segmenting lung lesions. The Universal Model \cite{clip} utilizes contrastive learning to integrate simple text-image information for multi-organ tasks.
Transformer-based models have further enhanced multi-modal integration. For example, LViT \cite{lvit} combines lesion location text input with a U-shaped Visual Transformer (ViT) to accurately segment pneumonia lesions. 
However, these methods often lack detailed morphological descriptions of lesions and involve high computational costs for text-image fusion.

Moreover, existing multi-modal approaches encounter limitations in detecting liver tumors, particularly in identifying small lesions and resolving blurred tumor boundaries. Traditional image processing and multi-scale analysis have shown utility; for instance, SBCNet \cite{sbcnet} employs Sobel operators to refine tumor boundary detection, while DA-Tran \cite{DA-Tran} applies 3D transformers for multi-scale segmentation on CT images, enhancing small lesion detection. However, these methods can inadvertently magnify non-target regions, leading to interference from adjacent organs.

\begin{figure*}[h]
    \centering
    \includegraphics[width=0.80\linewidth]{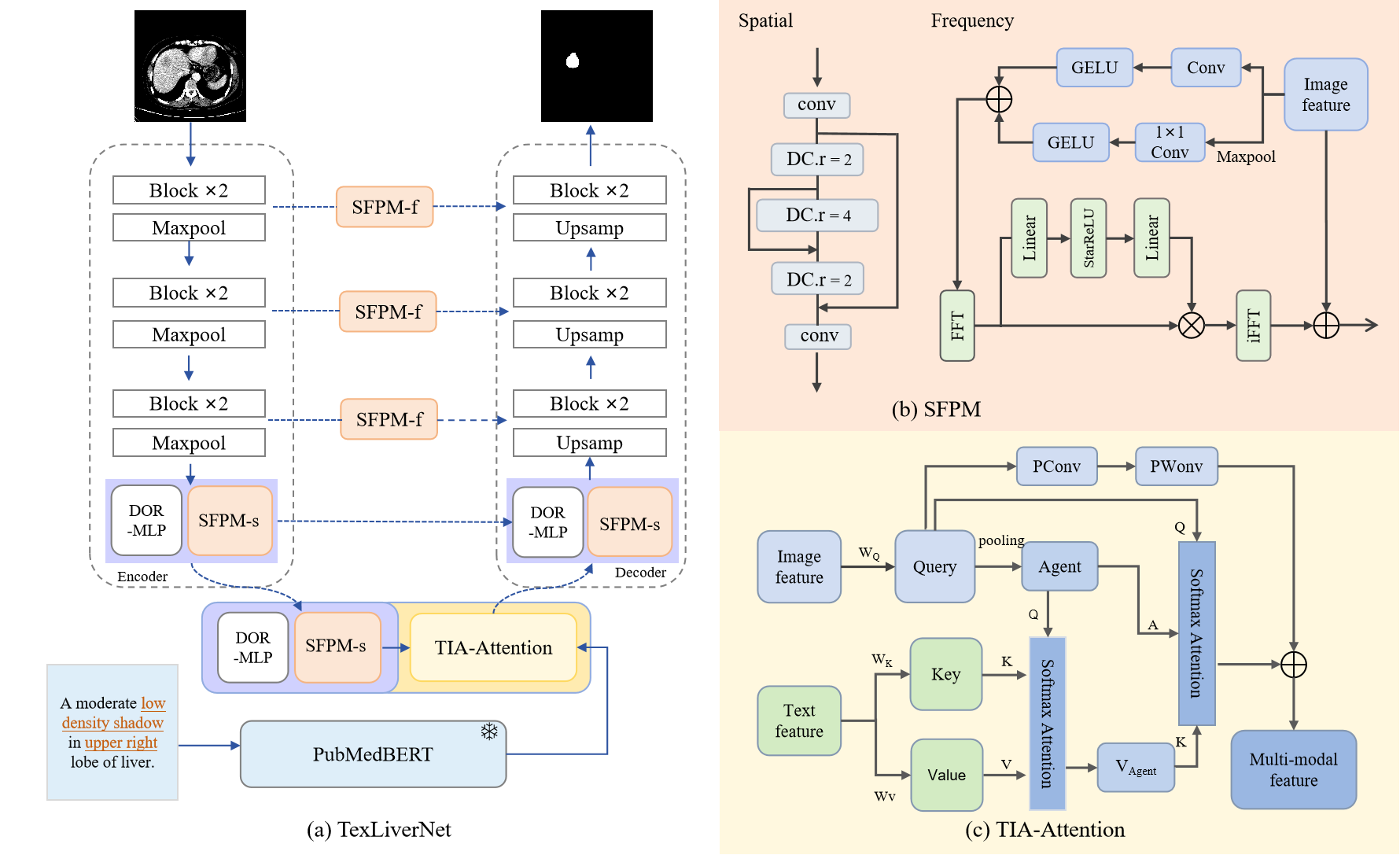}
    \caption{The architecture of the proposed TexLiverNet.}
    \label{fig:enter-label}
\end{figure*}

We argue that integrating domain-specific medical knowledge can effectively guide network learning. Hence, we developed the Visual-Text Liver Tumor Segmentation (VTLiTS) dataset, which provides both imaging data and more medically informed textual descriptions. 
We propose a novel model, TexLiverNet, leveraging spatial-frequency domain-enhanced imaging features alongside these text annotations. By efficiently fusing both modalities, TexLiverNet addresses challenges posed by small targets and unclear lesion boundaries, leading to more precise liver tumor segmentation.

Contributions are summarized as follows:
1) We proposed a text-image agent cross-attention module to enhance efficient interaction between the two modalities. The intermediate agent mechanism is adopted to reduce the computation complexity of the attention mechanism. 2) We improve texture detail preservation by integrating spatial and adaptive frequency domain perception capabilities, restoring deep information lost in images and clarifying lesion boundaries.
3) Experimental comparison with the state-of-the-art methods on two datasets demonstrate the effectiveness of TexLiverNet.

\section{method}
\label{sec:format}
The architecture of our proposed TSFNet is shown in Fig. 1(a), extending the encoder-decoder structure with two main components: TIA-attention for effective text-image interaction and SFPM for spatial and frequency domain texture perception. The initial encoder-decoder layers employ standard 3x3 convolutions, while the fourth and bottleneck layers utilize RollingUnet's DOR-MLP (Double Orthogonal Rolling-MLP) \cite{rolling} to capture long-range dependencies, enhanced by spatial awareness modules for finer texture details. Compared to Transformer-based architectures, this design significantly reduces computational complexity.

At the bottleneck, text features from a frozen PubMedBERT \cite{pubmedbert} are fused with image features through TIA-attention. These multi-modal features, combined with SFPM skip connection features, are then integrated within the decoder to enable precise tumor segmentation. The following sections provide further module details.
\subsection{Text-Image Agent Attention (TIA-Attention)}
\label{ssec:subhead}
The cross-attention mechanism effectively combines image and text information to enhance multi-modal task performance. However, traditional cross-attention methods are often computationally intensive.
To enable efficient fusion of these heterogeneous modalities with reduced computational overhead, we draw inspiration from the agent attention mechanism \cite{agent} and propose a text-image aware agent attention module in the network’s bottleneck layer, as illustrated in Fig.1(c).
Given the image features \( f_I \in {R}^{C \times H \times W} \) and text features \( f_T \in {R}^{L \times C} \), we first apply a \(1 \times 1\) convolution ($W_Q,W_K,W_V$) to linearly align the features, ensuring consistency in the number of channels. Next, we incorporate the corresponding learnable positional encoding ($P_I,P_T$) to generate the image query (\( Q \)), key (\( K \)), and value (\( V \)) representations. 
This process is formalized in Equ. 1:
\begin{align}
     \tiny Q \!&=\! f_I W_Q\!+\!P_I, K \!= \!f_T W_K\!+\!P_T,V \!=\! f_T W_V\!+\!P_T ~.
     \label{eq:shared}
\end{align}

To further optimize computational efficiency, we apply adaptive pooling to downsample the image query \( Q \), retaining essential context while reducing its spatial dimensions. This results in an image agent \( A \in {R}^{c \times h \times w} \), which performs softmax attention with \( K \) and \( V \) to compute the agent-conditional value matrix \( V_A \). Next, \( V_A \) is broadcast to a second softmax attention layer, acting as the key, along with the image query \( Q \) and agent \( A \), to compute the weight matrix (see Equ. 2). This agent-based process \( O^{Agent} \) bypasses the direct similarity calculation between \( Q \) and \( K \), significantly reducing computational cost.
\begin{align}
    &O^{Agent} =\operatorname{softmax}\left(\frac{Q A^{T}}{\sqrt{d}}\right) \operatorname{softmax}\left(\frac{A K^{T}}{\sqrt{d}}\right) V  ~.
\end{align}
where $d$ is the channel dimension of $F_I$ and $F_S$.

To further enrich image feature diversity, we introduce a lightweight diversity recovery module consisting of PConv \cite{pconv} and PWConv layers. PConv processes a portion of the image feature channels \( f_I \), while PWConv extracts information across all channels, minimizing memory access and preserving accuracy. The final multi-modal feature \( f_S \in {R}^{C \times H \times W} \) is obtained by adding the attention feature to the output of the recovery module. This process is shown in Equ. 3:
\begin{align}
    f_S = O^{Agent}(f_{I},f_{T})+(PWconv(PConv(f_{I} ))) ~.
\end{align}
\subsection{Spatial-Frequency Perception Module (SFPM)}
\label{ssec:subhead}
SFPM comprises two main components: spatial domain perception (SFPM-s) and adaptive frequency domain perception (SFPM-f).

{\bf SFPM-s.}
In deeper network layers, continuous downsampling can lead to a loss of fine image features. Although the DOR-MLP module effectively extracts global information, it lacks precision in capturing local details, which is critical for identifying small lesions. To address this, we incorporate the SFPM-s spatial perception module in a parallel structure within the network. The SFPM-s module comprises two standard convolutional layers and three dilated convolutional layers with dilation rates (DC.r) of 2, 4, and 2, respectively. Residual connections facilitate efficient inter-layer information transfer (see Fig.1(b)). Finally, the output of SFPM-s is combined with the DOR-MLP output, enhancing detail restoration in small lesion areas and providing comprehensive image features.

{\bf SFPM-f.}
In the image decoding stage, the network’s depth often causes texture detail loss, which skip connections can partially mitigate but may introduce noise. Since high-frequency components are essential for capturing fine structural details in medical images \cite{global}, we incorporate an adaptive frequency domain perception module (SFPM-f) alongside skip connections to enhance texture and clarify liver tumor boundaries. Specifically, we extract image features $I_{s}$ from the skip connections via max pooling and $1 \times 1$ convolution, then combine them with the original image features to obtain coarse high-frequency information, $H_S$, as shown in Fig.1(b).

The coarse high-frequency domain features undergo a fast Fourier transform (FFT), represented by $F$, and are then enhanced through an adaptive frequency domain filter, denoted as $G(\cdot)$. This adaptive frequency enhancement technique effectively distinguishes relevant high-frequency details from noise, minimizing noise amplification while precisely capturing fine details. This approach prevents false enhancement of noise and improves the delineation of various regions within the image. To maintain network efficiency, the adaptive frequency domain filter is designed with only two linear layers and optimized using the StarReLU nonlinear function. The formula for $G(\cdot)$ is provided in Equ. 4:
\begin{align}
    G(\cdot) &= Linear(StarRelu(Linear(\cdot))) ~.
\end{align}

Finally, SFPM-f is converted back to the spatial domain by fast Fourier inverse transform and fused with the image features to obtain frequency domain enhanced high frequency texture features $H$. Equ. 5 gives the above process.

\begin{align}
    H&=F^{-1} [G(F(H_{s}))*F(H_{s}) ]+I_{s} ~.      
\end{align}
We combine the spatial domain and frequency domain information to supplement the image information in an all-round way, effectively improving the network's extraction of texture details.

\begin{table*}[h]
    
    \centering
    \begin{tabular}{l|*{4}{c}|*{4}{c}}
    \toprule
    \multicolumn{1}{c|}{\multirow{2}{*}{Model}} & \multicolumn{4}{c|}{3Dircadb}&\multicolumn{4}{c}{VTLiTS}\\
    \cline{2-9}
    \multicolumn{1}{c|}{} & Dice(\%)$\uparrow$ &DG(\%)$\uparrow$ &VOE(\%)$\downarrow$&RAVD$\downarrow$ &Dice(\%)$\uparrow$ &DG(\%)$\uparrow$ &VOE(\%)$\downarrow$&RAVD$\downarrow$\\
    \midrule
    Unet\cite{unet}&63.12&74.07&48.40&0.3807&69.99&81.50&39.43&0.2114 \\
    Unet++\cite{unet++}&68.82&76.84&44.43&0.2290&76.84&85.79&34.65&0.0054 \\
    AttUnet\cite{attention}&67.57&75.79&45.65&0.3573&72.06&83.85&36.95&\bf0.0015  \\
    TransUnet\cite{transunet}&65.57&71.60&48.64&0.3936&69.33&76.70&42.23&0.1528  \\
    UCTransNet\cite{uctransnet}&65.76&72.20&47.95&0.3423&66.34&79.39&49.83&0.3828  \\
    RollingUnet-\textit{base} \cite{rolling}&73.11&81.01&38.28&0.0090&80.39&88.17&28.53&0.0079  \\
    \midrule
    TGANet\cite{tganet}&71.63&79.76&40.65&0.1305&76.83&86.91&32.58&0.1053  \\
    GLoRIA\cite{gloria}&71.20&78.36&41.06&0.0064&74.21&84.69&34.66&0.1536  \\
    LanGuideSeg\cite{ariadne}&73.91&82.65&37.95&0.1139&81.24&87.23&26.98&0.3256  \\
    LViT\cite{lvit}&72.08&80.97&38.93&0.0087&80.34&88.52&28.76&0.0036  \\
    \textit{Ours}-
    TexLiverNet&\bf75.26&\bf84.35&\bf36.74&\bf0.0062&\bf82.74&\bf90.33&\bf25.41&0.0027  \\
    \bottomrule
    \end{tabular}
    \caption{\textbf{Quantitative comparison with state-of-the-art on two liver tumor datasets}}
\end{table*}

\section{experiments}
\label{sec:pagestyle}
{\bf Dataset.}The evaluation utilizes two datasets: a private visual-text liver tumor segmentation dataset, VTLiTS, comprising 81 scans and 5,173 images, and the public dataset 3Dircadb \cite{3DIRCADB}, containing 20 scans and 574 images, to assess the model’s effectiveness in small-sample scenarios.
The dataset was divided into training and validation sets following an 8:2 random patient ratio. To complement image annotations, text annotations were further enriched to capture specific tumor characteristics such as size, location, and density morphology. For instance, a representative annotation might read, “a moderately low-density shadow in the upper right lobe of the liver with blurred margins”.
In VTLiTS, text annotations were initially extracted from diagnostic reports and subsequently reviewed and refined by clinical experts. In 3Dircadb, two experts in radiation oncology performed a detailed film-by-film text annotation expansion.

{\bf Experiment settings.} 
We convert the raw 3D data into 2D slices.
The Hounsfield intensity range of each slice is clipped to [-200, 200] and normalized to a [0,1] grayscale range. 
Each slice is resized to \(256 \times 256\) pixels, with data scaling for data augmentation. The loss function combines BCE and Dice loss (8:2 ratio). Training uses the AdamW optimizer with a weight decay of 1e-4, momentum of 0.9, and a cosine learning rate schedule (initial rate3e-4, minimum 1e-4). Batch size is 32, with a maximum of 200 epochs. 
All experiments are conducted on two NVIDIA GeForce RTX 3090 GPUs.

{\bf Evaluation metrics.} Four commonly used metrics for 2D liver tumor assessment were used to evaluate performance: Dice Similarity Coefficient (Dice), Dice Global Index (DG), Volume Overlap Error (VOE), and Relative Absolute Volume Difference (RAVD).

{\bf Quantitative evaluation.} We compared TexLiverNet with ten standard methods for single-modal image segmentation (upper section of Table 1) and multi-modal segmentation (lower section of Table 1). Overall, TSFNet achieved state-of-the-art segmentation performance on both the 3Dircadb and VTLiTS datasets. Compared to the best-performing single-modal method, RollingUnet, our approach improved the Dice score by 2.15\% and 2.38\%, the Dice Global score by 3.34\% and 2.16\%, and the VOE by 1.54\% and 3.12\%, respectively. 
Compared with the top text-image multi-modal method, LanGuideSeg, TexLiverNet demonstrated gains of 1.35\% and 1.5\% in Dice, 1.7\% and 3.1\% in Dice Global, and 1.21\% and 1.57\% in VOE, respectively.

\begin{table}[h]
    \centering
    \begin{tabular}{lccc}
    \toprule
    Networks&Dice (\%)$\uparrow$ &VOE (\%)$\downarrow$\\
    \midrule
    base&80.39&28.53  \\
    base+TIA-Attention&81.83&26.08  \\
    base+SFPM&81.36&25.92  \\
    base+TIA-Attention+SFPM&\bf82.74&\bf25.41  \\
    \bottomrule
    \end{tabular}
    \caption{\textbf{Ablation study of TexLiverNet on VTLiTS dataset.}}
\end{table}
{\bf Ablation study.} Table 2 lists the comprehensive ablation studies performed on the VTLiTS dataset to assess the contribution of each module. Starting with RollingUnet as a baseline, we progressively added textual information and spatial frequency perception components. Adding the SFPM and TIA-attention modules improves the comprehensiveness of the depth and image information, and when only text was added for TIA-attention fusion, the Dice score improved by 1.44\% and the VOE decreased by 2.45\%, demonstrating the important role of medical description. The segmentation performance is optimal when the two modules work jointly.

\begin{figure}[h]
    \centering
    \includegraphics[width=\linewidth]{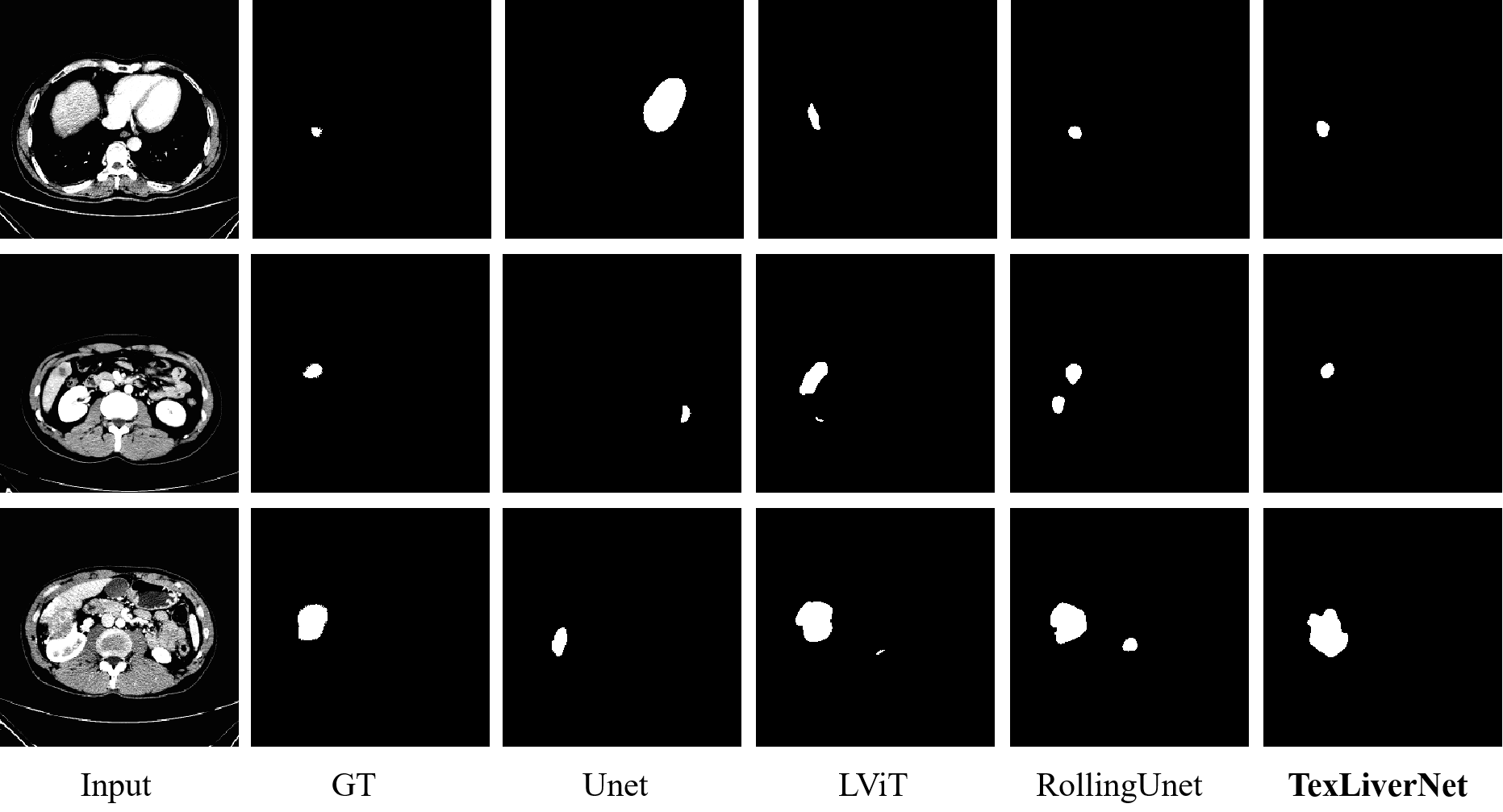}
    \caption{Qualitative comparison.}
    \label{fig:enter-label}
\end{figure}
{\bf Qualitative evaluation.} Fig.2 presents a qualitative comparison of segmentation results produced by our TSFNet, alongside several prominent methods on the 3Dircadb and VTLiTS datasets. For challenging cases, including low-contrast and small targets (row 1), small tumors (row 2), and blurred lesion boundaries (row 3), TSFNet demonstrates superior semantic segmentation capabilities, substantially reducing false segmentations. In contrast, Unet, RollingUnet, and LViT exhibit more prominent false segmentation in these scenarios. These results suggest that incorporating medical textual information and spatial domain frequency information into TexLiverNet's learning mechanism enables it to capture the textural details of lesions in finer detail, resulting in more accurate segmentation of lesion regions and shapes.
\section{conclusion}
\label{sec:print}
This paper introduces TexLiverNet, a liver tumor segmentation network designed to seamlessly integrate textual information with medical images. TexLiverNet employs an agent-based cross-attention module to achieve efficient, low-complexity fusion of text and image modalities. Additionally, it incorporates spatial and adaptive frequency domain perception to capture essential texture details, effectively merging global and local information to improve the detection of small tumors often overlooked in deep networks. Validation on two liver tumor datasets shows that TexLiverNet’s multimodal segmentation approach consistently surpasses traditional image-only methods. TexLiverNet offers a promising framework for early liver tumor segmentation, leveraging textual data to enhance diagnostic precision and efficiency in clinical applications.

\section{compliance with ethical standards}
\label{sec:page}

Informed consent was obtained from all individual participants involved in the study.

\section{ACKNOWLEDGMENTS}
\label{sec:fund}
This work was supported by the Science and Technology Program of Guangzhou, China
(202206010093). 
We thank Dr. Xiaojun Hu and Dr. Yingfang Fan from the Third Affiliated Hospital of the Southern Medical University and Zhijie Pan from the Department of Radiology, Affiliated General Hospital of Shanghai Jiao Tong University for their expertise and valuable contributions.

\bibliographystyle{IEEEbib}
\bibliography{strings,refs}

\end{document}